\documentclass[floatfix,aps,prl,twocolumn,showpacs]{revtex4}
\usepackage{mathrsfs}
\usepackage{graphicx}
\usepackage{amsmath}
\usepackage{amssymb}

\newcommand{\C}{{\mathcal{C}}}
\newcommand{\G}{{\mathcal{G}}}
\newcommand{\cGi}{{\mathcal{G}\backslash i}}

\begin{document}

\title{Ground-State Entropy of the Random Vertex-Cover Problem}

\author{Jie Zhou and Haijun Zhou}

\affiliation{Institute of Theoretical Physics, Chinese Academy of Sciences,
Beijing 100190, China}

\date{\today}

\begin{abstract}
Counting the number of ground states for a spin-glass or NP-complete
combinatorial
optimization problem is even more difficult than the already hard
task of finding a single ground state. In this paper the entropy of
minimum vertex-covers of random graphs is estimated through a set of
iterative equations based on the cavity method of 
statistical mechanics. During the iteration both the cavity entropy contributions
and cavity magnetizations for each vertex are updated.
This approach overcomes the difficulty of iterative divergence
encountered in the zero temperature
first-step replica-symmetry-breaking (1RSB) spin-glass theory.
It is still applicable when the 1RSB mean-field theory is no longer stable.
The method can be extended to compute the entropies of ground-states and
metastable minimal-energy states for other random-graph spin-glass systems.
\end{abstract}


\maketitle

%
%
%Introduction
%
%

A combinatorial optimization (CO) problem is defined by an energy
function $E(\vec{\sigma})$ on configurations $\vec{\sigma}
=(\sigma_1, \sigma_2, \ldots, \sigma_N)$ of a 
$N$-dimensional space, where each variable $\sigma_i$ has only a finite
number of states (e.g., $\sigma_i = \pm 1$).
For CO problems in the non-deterministic polynomial-complete (NP-complete)
class, searching for configurations of energy
equal or very close to the lowest possible value is in general
a very difficult task.
Statistical physicists relate this computational
hardness to the emergence of complex structures in the problem's configuration
space and the proliferation of metastable macroscopic states (macrostates).
The entropy spectrum (number of configurations at each
minimal-energy level) gives a characterization of the energy landscape
of a hard CO problem. 
Using the first-step replica-symmetry-breaking (1RSB)
cavity method of spin-glass theory
\cite{Mezard-Parisi-2001,Mezard-Parisi-2003,Mezard-etal-2002},
the ground-state energy densities for several
hard CO problems on random graphs have been
calculated with high precision (see, e.g.,
Refs.~\cite{Mezard-Parisi-2003,Weigt-Zhou-2006,Mezard-Tarzia-2007}). 
But estimating the ground-state entropy and the entropies of
metastable states is still a challenging theoretical and computational issue.

In the 1RSB cavity approach, the ground-state energy $E^0$ of
a hard CO problem is evaluated by first assuming the configuration space
of the system can be clustered into many macrostates.
A variable $i$ experiences in each macrostate an integer-valued field $h_i$.
The distribution of this field among all the
macrostates, $P_i ( h_i )$, is then obtained
 through an iterative numerical scheme
\cite{Mezard-Parisi-2003}.
To calculate the ground-state entropy, a conventional technique is
to introduce a temperature $T$ and expand the field $h_i$ to first
order in $T$ at the limit $T\rightarrow 0$ \cite{Zdeborova-Mezard-2006}:
\begin{equation}
	\label{eq:field-zero-limit}
	h_i = m_i + T  r_i \ ,
\end{equation}
with $m_i$ being an integer and $r_i$ a finite real value.
A set of iterative equations are derived to obtain the joint distribution
$P_i(m_i, r_i)$ of the values $m_i$ and $r_i$ for each variable $i$.
The ground-state entropy $S^0$ is then the first order term in
the free energy expansion $F(T)= E^0 - T S^0$.
This approach works
for some relatively simple problems 
(e.g., graph-matching whose configuration space
is ergodic \cite{Zdeborova-Mezard-2006},
random $Q$-coloring and $K$-SAT which have zero ground-state energy
\cite{Mezard-etal-2005,Krzakala-etal-PNAS-2007,Montanari-etal-2008})
but it fails for many other NP-complete CO problems, for which
the ground-state energy $E^0$ is positive and the $T\rightarrow 0$ 1RSB
cavity equations are not stable. 
For these later systems, at  $T=0$ the field $h_i$ is not necessarily
an integer and the correction $r_i$ in Eq.~(\ref{eq:field-zero-limit})
usually diverges at the $T\rightarrow 0$ limit. 
Consequently, although the $T\rightarrow 0$ 1RSB cavity method is able to
estimate the ground-state energy of a hard CO problem with high accuracy
(as only the value of $m_i$ but not that of $r_i$ is used),
it often reports a negative or divergent ground-state entropy.
 
In this paper we use a different way to estimate the ground-state
entropy of a CO problem or a finite-connectivity spin-glass. We work
directly at temperature zero and, within the 1RSB cavity framework, 
calculate both the cavity magnetizations and cavity entropies of each variable
in each macrostate. A very small cutoff $\epsilon$ is naturally
introduced in the iteration of cavity magnetizations to avoid divergence
in the iteration of cavity entropies.
This approach gives good results
when tested on the random vertex-cover problem, the random $3$-SAT problem,
and the random $\pm J$ spin-glass model, even when the $T\rightarrow 0$
1RSB mean-field spin-glass theory is no longer stable.
It is also able to
calculate the entropies for other minimal-energy levels. Here we focus on the
vertex-cover problem (a prototypical NP-hard problem \cite{Hartmann-Weigt-2003})
to demonstrate
the main ideas of this method. Detailed calculations on
other model systems will be reported elsewhere.

%
% RS formulas
%

Let us first briefly introduce the vertex-cover problem
\cite{Weigt-Hartmann-2000,Weigt-Hartmann-2001}.
A graph $\G$ contains $N$ vertices
and $M$ edges. Each edge $(i,j)$ is between
a pair of vertices $i, j$.
The mean connectivity of the graph is $c = 2 M / N$,
which is the number of edges a  vertex on average is connected to.
A vertex cover $C$ for graph $\G$ contains a subset of vertices
of $\G$ such that for each edge $(i,j) \in \G$, at least one of
its extremities is in $C$. If a vertex $i$ is contained in 
a vertex-cover $C$, we say it is covered in $C$.
The energy of a vertex-cover $C$ is defined as its cardinality,
$E(C)= |C|$. Graph
$\G$ has many vertex covers, with energy
ranging from the maximum value $N$ to the minimum value
$E^0_{\G}= \min_C |C|$.  We denote by $\C^0_{\G}$
the set of minimum vertex covers (MVCs) of graph $\G$:
$\C^0_{\G} = \{ C : |C|=E^0_{\G} \}$.
To determine exactly the energy $E^0$ of MVCs for a graph
in general is a very hard computational problem;
to count the number of MVCs is even harder. Using the cavity method
of statistical mechanics, the mean energy density of MVCs for
random graphs has been evaluated in
Refs.~\cite{Weigt-Hartmann-2000,Zhou-2005a,Weigt-Zhou-2006}.
The entropy of MVCs for a random graph was also estimated
by computer simulations \cite{Weigt-Hartmann-2001}.
This work complements Ref.~\cite{Weigt-Hartmann-2001} by giving
an analytical estimation of the ground-state entropy $S^0_{\G} \equiv
\log |\C^0_{\G} |$.

We denote by $\pi_i$ the probability of a vertex $i$ being contained in the MVCs,
\begin{equation}
	\pi_i = \frac{1}{ | \C^0_{\G} |}
	\sum\limits_{ C \in \C^0_\G} \mathbb{I} ( i \in C) \ ,
	\label{eq:pi}
\end{equation}
where $\mathbb{I}$ is the indicator function.
Similar to $\pi_i$, for a vertex $j\in \partial i$ (the set of vertices which
are connected to $i$ by an edge),  we denote by  
$\pi_{j|i}$ its probability of being in the MVCs of the
cavity graph $\cGi$ (which is obtained from $\G$ by removing vertex $i$ and all its edges).

We consider here the ensemble of random graphs with mean connectivity
$c$. When the graph size $N$ is sufficiently large the length of a
loop in the graph is of order $\log_c(N)$, and a random graph $\G$ is
locally tree-like.
Consider the set $\partial i$ of vertices in the neighborhood
of a randomly chosen vertex $i$. If the edges between $i$ and these
vertices are deleted, in the cavity graph $\cGi$
the shortest length between any two vertices $j, k \in \partial i$
diverges logarithmically with $N$. As a first step, it is therefore assumed
that these vertices are uncorrelated in the cavity
graph $\cGi$, and the probability of finding these
vertices in a MVC of $\cGi$ can be expressed in
a factorized form:
\begin{equation}
	\label{eq:factorization-assumption}
	\mathbb{P}\bigl( \partial i \subset  C: 
	C \in \C^0_{\cGi} \bigr)
	=  \prod\limits_{j\in i} \pi_{j|i} \ .
\end{equation}
Equation (\ref{eq:factorization-assumption}) is called the Bethe-Peierls
approximation and the corresponding cavity method is referred to be
replica-symmetric (RS). Under this assumption, the energy and entropy of
MVCs for a random graph $\G$ can be expressed as
\cite{Weigt-Zhou-2006}
\begin{eqnarray}
	E^0_\G  &= & \sum\limits_{i} \Delta E_{i}
	- \sum\limits_{(i,j)} \Delta E_{(i,j)}
	\ , 
	\label{eq:energy} \\ 
	S^0_\G  & = & \sum\limits_{i} \Delta S_{i}
	-\sum\limits_{(i,j)} \Delta S_{(i,j)}
	\ , 
	\label{eq:entropy}
\end{eqnarray}
where $\Delta E_x$ and $\Delta S_x$ are, respectively, the
contribution to the system's ground-state energy and entropy from
an vertex ($x=i$) or an edge ($x=(i,j)$). For example,
$\Delta E_i = E^0_\G- E^0_{\cGi}$ and
$\Delta S_i = \log |\C^0_\G | - \log | \C^0_{\cGi} |$.

We shall distinguish three situations when
writing down the expressions for $\Delta E_i$ and $\Delta S_i$.
The total number of MVCs of cavity graph $\cGi$
which contain the set $\partial i$ is
$|\C^0_{\cGi}| 
\prod_{j \in \partial i} \pi_{j|i}$. This number is positive if
all the cavity probabilities $\pi_{j|i} >0$.
In this case, when vertex $i$ is added, it will not be in
any MVC of $\G$ but the set $\partial i$
will be contained in every MVC of $\G$.
Then we have $\pi_{i}=0$ (vertex $i$ is always non-covered),
 $\Delta E_i=0$, and 
$| \C^0_\G|  = | \C^0_{\cGi} | 	\prod_{j\in \partial i} \pi_{j|i}$,
and the change in entropy is
$\Delta S_i =\sum_{j\in i} \log \pi_{j|i}$.
On the other hand, if in the set $\partial i$ there are at least two vertices
which are not in any MVCs of cavity
graph $\cGi$, then vertex $i$ will be present in all the MVCs of graph $\G$.
Adding vertex $i$ to a MVC of cavity graph $\cGi$ results in a MVC of
graph $\G$ ($| \C^0_\G| = | \C^0_{\cGi} |$).
In this case $\pi_i=1$ ($i$ is always covered),
 $\Delta E_i = 1$, and $\Delta S_i = 0$.

Now we consider the remaining case, namely in
the set $\partial i$ there is only one vertex (say 
$j$) which is not in any MVC of graph $\cGi$. In this case, each
MVC of graph $\G$ contains either vertex $i$ or vertex $j$ but not both.
The energy increase is $\Delta E_i = 1$, and the total
number of MVCs for graph $\G$ is
\begin{equation}
	\label{eq:Number-unfrozen}
	|\C^0_\G| = 
	|\C^0_{\cGi}| +
	|\C^{0}_{\G\backslash i,j}|
	\prod\limits_{k\in \partial i \backslash j} \pi_{k|i}
	\ ,
\end{equation}
where $\C^{0}_{\G\backslash i,j}$ is the
complete set of MVCs for the cavity graph $\G\backslash i, j$ (the
remaining graph after further removing vertex $j$ from $\cGi$),
and  $\partial i \backslash j$ is the set of direct neighbors
except $j$ of vertex $i$.
The first term on the right hand side
of the equality in Eq.~(\ref{eq:Number-unfrozen})
 is the number of MVCs which contain vertex $i$, 
while the second term is the number of MVCs which contain $j$.
In this case vertex $i$ is said to be unfrozen.
The entropy change
$\Delta S_i = \log\bigr( 1+ e^{-\Delta S_{j|i}}
\prod_{k\in \partial i \backslash j} \pi_{k|i} \bigr)$, and 
$\pi_{i}= 1/\bigl(1+e^{-\Delta S_{j|i}}
\prod_{k\in \partial i \backslash j} \pi_{k|i} \bigr)$, where
$ \Delta S_{j|i} = \log |\C^{0}_{\cGi}| -
 \log | \C^{0}_{\G\backslash i, j}|$ is the entropy change due to
adding vertex $j$ to the cavity graph $\G\backslash i, j$. We
refer $\Delta S_{j|i}$ as a cavity entropy of vertex $j$. 
Notice that in this case, to properly calculate $\Delta S_i$, one needs not
only to consider all the MVCs of cavity graph $\cGi$
but also vertex-covers with a higher energy $E^0_{\cGi}+1$. 

The energy and entropy contribution of an edge $(i,j)$ can also be
obtained similarly. If at least one of the cavity probabilities
$\pi_{j|i}$ and $\pi_{i|j}$ is
positive, then $\Delta E_{(i,j)}=0$ and $\Delta S_{(i,j)}=\log\bigl(
1-(1-\pi_{i|j}) (1-\pi_{j|i}) \bigr)$. On the other hand,
if both $\pi_{i|j}$
and $\pi_{j|i}$ are zero, then
$\Delta E_{(i,j)}=1$ and $\Delta S_{(i,j)}=\log(e^{-\Delta S_{i|j}} +
e^{-\Delta S_{j|i}})$.

\begin{figure}
\includegraphics[width=0.8\linewidth]{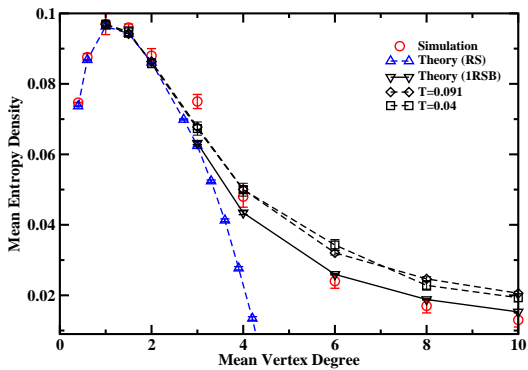}
\caption{\label{fig:entropy}
(Color Online) Ground-state entropy density of the vertex-cover problem as a
function of the mean vertex degree $c$ of the random graph. Circles are
simulation data \cite{Weigt-Hartmann-2000}, up- and
down-triangles are, respectively,
theoretical results obtained by the replica-symmetric and the
1RSB cavity method. The finite-temperature entropy density values at $T=0.091$ and
$0.04$ are also shown as a comparison.
}
\end{figure}

On each of the $2 M$  directed edges $j\rightarrow i$ of graph $\G$ there is
a cavity probability $\pi_{j|i}$ and a cavity entropy 
$\Delta S_{j|i}$. After all these $4 M$ values are determined,
the total ground-state energy and entropy can then be
evaluated  using Eqs.~(\ref{eq:energy}) and (\ref{eq:entropy}).
Starting from a random initial condition, 
the values of $\pi_{j|i}$, $\Delta S_{j|i}$, and the cavity energy
increase $\Delta E_{j|i}$ can be determined by
the following iterative algorithm. Consider vertices $k$ in the neighborhood
$\partial j\backslash i$ of a vertex $j$:
\begin{enumerate}
\item[(1)] If all $\pi_{k|j} >0$, then $\pi_{j|i}=0$, $\Delta E_{j|i}=0$, 
$\Delta S_{j|i}= \sum_{k \in \partial j\backslash i} \log \pi_{k|j}$.
\item[(2)] If  two  or more $\pi_{k|j}=0$, then
$\pi_{j|i}=1$, $\Delta E_{j|i}=1$,  $\Delta S_{j|i}=0$.
\item[(3)] If only one $\pi_{k|j}=0$, then $\Delta E_{j|i}=1$, and 
	\begin{eqnarray}
	\pi_{j|i} & = & \Bigl(1+e^{-\Delta S_{k|j}}
	\prod_{l\in \partial j \backslash i,k} \pi_{l|j} \Bigr)^{-1} \ ,
	\label{eq:pi-ji} \\
	\Delta S_{j|i} & =  & \log\Bigl(1+ e^{-\Delta S_{k|j}}
	\prod_{l\in \partial j \backslash i,k} \pi_{l|j} \Bigr)  \ .
	\end{eqnarray}
\end{enumerate}
When using Eq.~(\ref{eq:pi-ji}) to update the cavity
probability $\pi_{j|i}$, sometimes $\pi_{j|i}$ becomes extremely close to
zero. A cutoff $\epsilon$ is therefore introduced in the numerical
scheme: If the obtained value of $\pi_{j|i}$ is less than $\epsilon$ we
set $\pi_{j|i}=0$. The theoretical entropy values shown in
Fig.~\ref{fig:entropy}
are obtained by setting a cutoff value of $\epsilon = 10^{-10}$
(more discussions on the choice of $\epsilon$ are mentioned below).

The above discussion concerns with one large graph $\G$. As we are
interested in the graph-averaged values for the ground-state
energy and entropy, a population dynamics simulation can be performed
similarly by storing a large array of $\pi_{j|i}$ and $\Delta S_{j|i}$
and then updating this array \cite{Mezard-Parisi-2001,Weigt-Zhou-2006}.
The ground-state entropy density $s^0 = \lim_{N\rightarrow \infty} S_\G^0/N$
as a function of mean connectivity $c$ of the random graph $\G$ is
evaluated by this numerical scheme (Fig.~\ref{fig:entropy}, up-triangles).
The ground-state entropy first increases with $c$ and
reaches a maximum at $c \approx 1$, then it decreases with $c$.
For $c < 2.7183$ the RS cavity method is known to be valid
\cite{Zhou-2005a,Weigt-Zhou-2006} and it correctly predicts the ground-state
entropy and energy density for the system. For $c > 2.7183$ however, the
RS prediction is systematically lower
than the simulation result of Hartmann and Weigt (circles in
Fig.~\ref{fig:entropy}) and even becomes negative for $c > 4.4$
(such an entropy crisis was also observed in the hitting set problem
\cite{Mezard-Tarzia-2007}). For $c>2.7183$ the entropy predicted by
the RS cavity method depends strongly on the cutoff $\epsilon$. If
a smaller $\epsilon$ value is used, the 
entropy decreases even faster with the mean connectivity
$c$.

%
% 1RSB situation
%

When the mean connectivity $c$ of the random graph is larger than
$2.7183$, the Bethe-Peierls approximation
Eq.~(\ref{eq:factorization-assumption}) is no longer a good assumption,
as there are strong long-range correlations among distant vertices
\cite{Zhou-2005a}. For example, consider two vertices
$i$ and $j$ whose
shortest-path length is of order $\log_c N$ and suppose these two vertices
both are unfrozen ($\pi_i > 0,  \pi_j>0$) among the space of MVCs of
graph $\G$. The Bethe-Peierls approximation assumes that the probability
of finding a MVC which contains both $i$ and $j$ is equal to $\pi_i \pi_j > 0$.
However, it may be the case that there is not a single MVC in which both
$i$ and $j$ are present
\cite{Zhou-2005a}. To take into account such long-range
correlations, in the 1RSB cavity theory the MVC set $\C^{0}_\G$ of a graph $\G$
is clustered into
many subsets $\C^{0, \alpha}_\G$ which are indexed by an index $\alpha$.
In each such subset it is assumed that the Bethe-Peierls approximation
still holds, and Eq.~(\ref{eq:factorization-assumption}) is replaced by
\begin{equation}
	\label{eq:factorization-assumption-1RSB}
	\mathbb{P}\bigl( \partial i \subset  C: 
	C \in \C^{0,\alpha}_{\cGi} \bigr)
	=  \prod\limits_{j\in i} \pi_{j|i}^\alpha \ .
\end{equation}
where $\pi_{j|i}^\alpha$ is the probability of vertex $j$ being in the
MVCs of the $\alpha$-th subset $\C^{0,\alpha}_\cGi$ of the cavity graph
$\cGi$. Because of Eq.~(\ref{eq:factorization-assumption-1RSB}), the
iterative equations for $\pi_{j|i}$ and $\Delta S_{j|i}$ as mentioned 
before are still valid in each subset of MVCs. To characterize
the property of different clusters, a  probability distribution 
$\mathcal{P}_{j|i}( \pi , s )$ is introduced on each directed edge
of the graph, which is equal to
the fraction of MVC clusters
with $\pi_{j|i}=\pi$ and $\Delta S_{j|i}=s$. For a single
graph $\G$ these $2 M$ probability
distributions $\mathcal{P}_{j|i}$ can again be obtained
by iterations, and the corresponding distribution of 
$\mathcal{P}_{j|i}(\pi, s)$ among all the edges of the random
graph can be obtained by mean-field population dynamics.

The iteration equation for $\mathcal{P}_{j|i}(\pi, s)$ reads
\begin{eqnarray}
 \mathcal{P}_{j|i}(\pi, s)  & \propto &  \int \prod\limits_{k\in \partial j \backslash i}
	\int {\rm d} \pi_{k|j} {\rm d} s_{k|j} \mathcal{P}_{k|j}(\pi_{k|j},  s_{k|j}) \nonumber \\
	& & e^{-y \Delta E_{j|i}} \delta(\pi-\pi_{j|i}) \delta (s- \Delta S_{j|i}) \ .
 \label{eq:1RSB-iteration}
\end{eqnarray}
The value of the re-weighting parameter $y$ in the above equation is chosen
such that a complexity parameter is equal to zero \cite{Mezard-Parisi-2003}.
In writing down Eq.~(\ref{eq:1RSB-iteration}), it is further assumed that the joint probability of observing the
cavity values $(\pi_{k|j}, \Delta S_{k|j}), (\pi_{l|j}, \Delta S_{l|j}),
\ldots$ for vertices $k, l, \ldots \in \partial j \backslash i$ can be
written in a factorized form:
\begin{equation}
	\label{eq:factorization-assumption-1RSB-2}
	\mathbb{P}\bigl(\pi_{k|j}, \Delta S_{k|j}; 
	\ldots \bigr)
 	= \prod\limits_{ m\in \partial j \backslash i}
 \mathcal{P}_{m|j}(\pi_{m|j}, \Delta S_{m|j}) \ .
\end{equation}
Details of the 1RSB numerical iteration scheme for the vertex-cover
problem can be found in Ref.~\cite{Weigt-Zhou-2006}.

\begin{figure}
\includegraphics[width=0.8\linewidth]{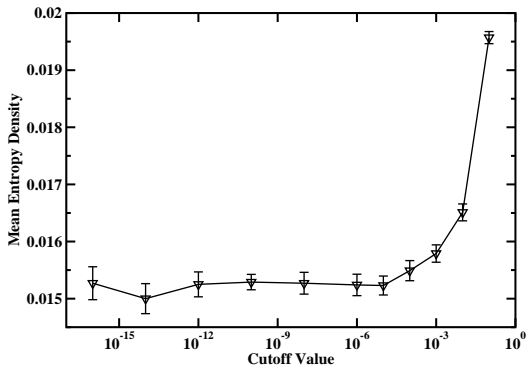}
\caption{\label{fig:cutoff}
Predicted mean ground-state entropy of the random vertex-cover problem
at connectivity $c=10$ as a function of the cutoff parameter $\epsilon$.
}
\end{figure}

The mean entropy density of 
MVCs for random graphs of mean connectivity $c$ as obtained by this 1RSB
cavity method is shown in Fig.~\ref{fig:entropy} (down-triangles).
The theoretical predictions are in good agreement with
simulation results \cite{Weigt-Hartmann-2001}. The mean ground-state
energy density as obtained by the present method is also in good agreement
with the  simulation and theoretical results of Ref.~\cite{Weigt-Zhou-2006}.
In the calculation we have used a cutoff value $\epsilon=10^{-10}$. 
Such a cutoff is necessary for the
random vertex-cover problem, as the 1RSB cavity approach is not stable to
further steps of replica-symmetry-breaking \cite{Zhou-etal-2007}.
Figure~\ref{fig:cutoff} shows that, the
predicted value of the mean ground-state
entropy density is not sensitive to the
cutoff parameter  $\epsilon$ when $\epsilon \leq 10^{-5}$. 

We also carry out lengthy population dynamics simulations 
at finite temperatures using different protocols.
The entropy values obtained at $T=0.091$ and $T=0.04$
are shown in Fig.~\ref{fig:entropy}. At these low temperatures although the obtained
energy density values are almost indiscernible from the ground-state values,
there is still a gap between the finite-temperature and the ground-state
entropy density when $c>2.7183$. If further lowering the temperature, the quality of the
simulation results deteriorate, possibly because of insufficient population size and insufficient
equilibrium and sampling times. We were also unable to remove this gap by  
using instead the expansion Eq.~(\ref{eq:field-zero-limit}), because the population dynamics diverges.
In comparison with these, the zero-temperature direct method is computationally much
efficient and also easier to implement.

%
% Conclusion
%

In summary, we have calculated the ground-state entropy for the random
vertex-cover problem using the 1RSB cavity approach of spin-glass theory.
In our method, both the cavity probabilities $\pi_{j|i}$ and cavity
entropies $\Delta S_{j|i}$ of each vertex $j$ in a cluster of MVC solutions
are recorded. 
We have paid special attention on unfrozen vertices (each of which belongs to
some but not all MVCs of the graph).
As demonstrated by Eq.~(\ref{eq:Number-unfrozen}),
the entropy contribution of an unfrozen $i$ comes not only from
the MVCs of the graph $\cGi$ but also form other higher-energy
configurations of $\cGi$. Similarly the cavity entropy $\Delta S_{j|i}$ of
a vertex $j$ also has two sources of contributions.
Equation~(\ref{eq:Number-unfrozen}) is rather simple for the vertex-cover
problem, while for some other NP-hard CO problems and spin-glass models (e.g.,
the random-graph $\pm J$ spin-glass) counting the entropy contribution of
an unfrozen vertex can be more complicated. 
A cutoff parameter $\epsilon$ is introduced so that if
$\pi_{j|i} < \epsilon$ in one cluster, it is set to be zero. 
With this cutoff parameter, the present cavity method can still give
good estimations on the ground-state entropy of a hard CO problem
or spin-glass system even if more steps of replica-symmetry-breaking are
needed to fully describe the system.

%
% Acknowledgement
%

We thank Alexander Hartmann and  Martin Weigt for sharing
their simulation data, Pan Zhang for helpful discussions,
and KITPC (Beijing) and NORDITA (Stockholm) for hospitality.
This work was partially supported by NSFC (Grant No. 10774150).

%\bibliography{/cygdrive/d/ResearchPapers/references}

\begin{thebibliography}{14}
\expandafter\ifx\csname natexlab\endcsname\relax\def\natexlab#1{#1}\fi
\expandafter\ifx\csname bibnamefont\endcsname\relax
  \def\bibnamefont#1{#1}\fi
\expandafter\ifx\csname bibfnamefont\endcsname\relax
  \def\bibfnamefont#1{#1}\fi
\expandafter\ifx\csname citenamefont\endcsname\relax
  \def\citenamefont#1{#1}\fi
\expandafter\ifx\csname url\endcsname\relax
  \def\url#1{\texttt{#1}}\fi
\expandafter\ifx\csname urlprefix\endcsname\relax\def\urlprefix{URL }\fi
\providecommand{\bibinfo}[2]{#2}
\providecommand{\eprint}[2][]{\url{#2}}

\bibitem[{\citenamefont{M{\'{e}}zard and Parisi}(2001)}]{Mezard-Parisi-2001}
\bibinfo{author}{\bibfnamefont{M.}~\bibnamefont{M{\'{e}}zard}}
  \bibnamefont{and} \bibinfo{author}{\bibfnamefont{G.}~\bibnamefont{Parisi}},
  \bibinfo{journal}{Eur. Phys. J. B} \textbf{\bibinfo{volume}{20}},
  \bibinfo{pages}{217} (\bibinfo{year}{2001}).

\bibitem[{\citenamefont{M{\'{e}}zard and Parisi}(2003)}]{Mezard-Parisi-2003}
\bibinfo{author}{\bibfnamefont{M.}~\bibnamefont{M{\'{e}}zard}}
  \bibnamefont{and} \bibinfo{author}{\bibfnamefont{G.}~\bibnamefont{Parisi}},
  \bibinfo{journal}{J. Stat. Phys.} \textbf{\bibinfo{volume}{111}},
  \bibinfo{pages}{1} (\bibinfo{year}{2003}).

\bibitem[{\citenamefont{M{\'{e}}zard et~al.}(2002)\citenamefont{M{\'{e}}zard,
  Parisi, and Zecchina}}]{Mezard-etal-2002}
\bibinfo{author}{\bibfnamefont{M.}~\bibnamefont{M{\'{e}}zard}},
  \bibinfo{author}{\bibfnamefont{G.}~\bibnamefont{Parisi}}, \bibnamefont{and}
  \bibinfo{author}{\bibfnamefont{R.}~\bibnamefont{Zecchina}},
  \bibinfo{journal}{Science} \textbf{\bibinfo{volume}{297}},
  \bibinfo{pages}{812} (\bibinfo{year}{2002}).

\bibitem[{\citenamefont{Weigt and Zhou}(2006)}]{Weigt-Zhou-2006}
\bibinfo{author}{\bibfnamefont{M.}~\bibnamefont{Weigt}} \bibnamefont{and}
  \bibinfo{author}{\bibfnamefont{H.}~\bibnamefont{Zhou}},
  \bibinfo{journal}{Phys. Rev. E} \textbf{\bibinfo{volume}{74}},
  \bibinfo{pages}{046110} (\bibinfo{year}{2006}).

\bibitem[{\citenamefont{M{\'{e}}zard and Tarzia}(2007)}]{Mezard-Tarzia-2007}
\bibinfo{author}{\bibfnamefont{M.}~\bibnamefont{M{\'{e}}zard}}
  \bibnamefont{and} \bibinfo{author}{\bibfnamefont{M.}~\bibnamefont{Tarzia}},
  \bibinfo{journal}{Phys. Rev. E} \textbf{\bibinfo{volume}{76}},
  \bibinfo{pages}{041124} (\bibinfo{year}{2007}).

\bibitem[{\citenamefont{Zdeborov{\'{a}} and
  M{\'{e}}zard}(2006)}]{Zdeborova-Mezard-2006}
\bibinfo{author}{\bibfnamefont{L.}~\bibnamefont{Zdeborov{\'{a}}}}
  \bibnamefont{and}
  \bibinfo{author}{\bibfnamefont{M.}~\bibnamefont{M{\'{e}}zard}},
  \bibinfo{journal}{J. Stat. Mech.: Theo. Exp.}, \bibinfo{pages}{P05003}
  (\bibinfo{year}{2006}).

\bibitem[{\citenamefont{M{\'{e}}zard et~al.}(2005)\citenamefont{M{\'{e}}zard,
  Palassini, and Rivoire}}]{Mezard-etal-2005}
\bibinfo{author}{\bibfnamefont{M.}~\bibnamefont{M{\'{e}}zard}},
  \bibinfo{author}{\bibfnamefont{M.}~\bibnamefont{Palassini}},
  \bibnamefont{and} \bibinfo{author}{\bibfnamefont{O.}~\bibnamefont{Rivoire}},
  \bibinfo{journal}{Phys. Rev. Lett.} \textbf{\bibinfo{volume}{95}},
  \bibinfo{pages}{200202} (\bibinfo{year}{2005}).

\bibitem[{\citenamefont{Krzakala et~al.}(2007)\citenamefont{Krzakala,
  Montanari, {Ricci-Tersenghi}, Semerjian, and
  Zdeborova}}]{Krzakala-etal-PNAS-2007}
\bibinfo{author}{\bibfnamefont{F.}~\bibnamefont{Krzakala}},
  \bibinfo{author}{\bibfnamefont{A.}~\bibnamefont{Montanari}},
  \bibinfo{author}{\bibfnamefont{F.}~\bibnamefont{{Ricci-Tersenghi}}},
  \bibinfo{author}{\bibfnamefont{G.}~\bibnamefont{Semerjian}},
  \bibnamefont{and}
  \bibinfo{author}{\bibfnamefont{L.}~\bibnamefont{Zdeborova}},
  \bibinfo{journal}{Proc. Natl. Acad. Sci. USA} \textbf{\bibinfo{volume}{104}},
  \bibinfo{pages}{10318} (\bibinfo{year}{2007}).

\bibitem[{\citenamefont{Montanari et~al.}(2008)\citenamefont{Montanari,
  {Ricci-Tersenghi}, and Semerjian}}]{Montanari-etal-2008}
\bibinfo{author}{\bibfnamefont{A.}~\bibnamefont{Montanari}},
  \bibinfo{author}{\bibfnamefont{F.}~\bibnamefont{{Ricci-Tersenghi}}},
  \bibnamefont{and}
  \bibinfo{author}{\bibfnamefont{G.}~\bibnamefont{Semerjian}},
  \bibinfo{journal}{J. Stat. Mech.: Theor. Exper.},  \bibinfo{pages}{P04004}
  (\bibinfo{year}{2008}).

\bibitem[{\citenamefont{Hartmann and Weigt}(2003)}]{Hartmann-Weigt-2003}
\bibinfo{author}{\bibfnamefont{A.~K.} \bibnamefont{Hartmann}} \bibnamefont{and}
  \bibinfo{author}{\bibfnamefont{M.}~\bibnamefont{Weigt}}, \bibinfo{journal}{J.
  Phys. A: Math. Gen.} \textbf{\bibinfo{volume}{36}}, \bibinfo{pages}{11069}
  (\bibinfo{year}{2003}).

\bibitem[{\citenamefont{Weigt and Hartmann}(2000)}]{Weigt-Hartmann-2000}
\bibinfo{author}{\bibfnamefont{M.}~\bibnamefont{Weigt}} \bibnamefont{and}
  \bibinfo{author}{\bibfnamefont{A.~K.} \bibnamefont{Hartmann}},
  \bibinfo{journal}{Phys. Rev. Lett.} \textbf{\bibinfo{volume}{84}},
  \bibinfo{pages}{6118} (\bibinfo{year}{2000}).

\bibitem[{\citenamefont{Weigt and Hartmann}(2001)}]{Weigt-Hartmann-2001}
\bibinfo{author}{\bibfnamefont{M.}~\bibnamefont{Weigt}} \bibnamefont{and}
  \bibinfo{author}{\bibfnamefont{A.~K.} \bibnamefont{Hartmann}},
  \bibinfo{journal}{Phys. Rev. E} \textbf{\bibinfo{volume}{63}},
  \bibinfo{pages}{056127} (\bibinfo{year}{2001}).

\bibitem[{\citenamefont{Zhou}(2005)}]{Zhou-2005a}
\bibinfo{author}{\bibfnamefont{H.}~\bibnamefont{Zhou}}, \bibinfo{journal}{Phys.
  Rev. Lett.} \textbf{\bibinfo{volume}{94}}, \bibinfo{pages}{217203}
  (\bibinfo{year}{2005}).

\bibitem[{\citenamefont{Zhou et~al.}(2007)\citenamefont{Zhou, Ma, and
  Zhou}}]{Zhou-etal-2007}
\bibinfo{author}{\bibfnamefont{J.}~\bibnamefont{Zhou}},
  \bibinfo{author}{\bibfnamefont{H.}~\bibnamefont{Ma}}, \bibnamefont{and}
  \bibinfo{author}{\bibfnamefont{H.}~\bibnamefont{Zhou}}, \bibinfo{journal}{J.
  Stat. Mech.: Theor. Exp.}, \bibinfo{pages}{L06001} (\bibinfo{year}{2007}).

\end{thebibliography}

\end{document}